# Short range attraction between two similarly charged silica surfaces

Ohad Zohar, Ilya Leizerson, Uri Sivan

Faculty of Physics and the Russell Berrie Nanotechnology Institute,

Technion – Israel Institute of Technology, Haifa 32000, Israel

November 24$^{th}$ 2005

**Abstract**

**Using an Atomic Force Microscope (AFM) we measure the interaction between two identically charged silica surfaces in the presence of saline solution. For pure NaCl the interaction is always repulsive. Upon addition of cobalt hexamine ions, $Co(NH_3)_6^{+3}$, the repulsion is gradually suppressed and a pronounced attraction develops at distances much shorter than the screening length. Higher concentrations of cobalt hexamine turn the attraction back into repulsion. Measurements of surface charge renormalization by the three valent cations provide their surface density and their association constant to the negatively charged silica surface. These estimates exclude interaction between two condensed Wigner crystals as an explanation for the attraction.**

High valence ions are known to induce aggregation and sedimentation in solutions containing charged macromolecules[1]. DNA condensation by $Z = 3$ and 4 cations, for instance, has been studied extensively in recent years. Other examples include aggregation of charged fibers, membranes, proteins, polymers, and colloids. A growing body of evidence point to the key role played by Coulomb correlations between the ions but surprisingly little is known experimentally on the force acting between surfaces in a regime where such correlations matter. The experiment described below was set to measure just that. It was motivated by scientific curiosity as well as by the broad range of applications of such attraction, from water waste treatment to gene delivery,

It has been argued that the correlations between large Z ions lead to their condensation into low dimensional Wigner crystals on charged surfaces[2-4], an assertion supported by the pronounced charge ordering found in numerical simulations[5]. Such condensation is consistent with the neutralization and reversal of surface charge by the added ions as evident, e.g., in Zeta potential measurements. Aggregation and charge neutralization occur simultaneously and they are therefore likely to be related. A particularly transparent model in this spirit has been proposed in Ref. 4. According to that reference, attraction results from interaction between the condensed Wigner crystals on the two plates. A charge on one plate and the oppositely charged correlation hole on the other plate attract each other. One of the important conclusions from our research is the ruling out of this explanation for attraction (see below).

A recent publication[6] proposes yet another mechanism for the attraction between two likely charged objects, provided they are not identical. Such asymmetry may result, for instance, from surface charge variation in the relevant interaction area. The attraction relies in this case on asymmetric charge condensation on the two surfaces and an induced polarization. The predictions of this model disagree again with our experimental findings (see below).

Overall, existing models do not account for the force curves presented below. Yet, the data provide indispensable information about a truly generic phenomenon, correlation induced attraction between identically charged bodies.

Attraction between macroscopically similar surfaces has been reported in the literature. For instance, the long range attraction between hydrophobic surfactant-coated surfaces[7] has recently been attributed to charge inhomogeneity[8]. Charge inversion due to over-screening by high valence ions may also lead to attraction between **dissimilar** surfaces. As the large Z-ion concentration is increased, one of the surfaces reverses its polarity before the other surface does, leading to "conventional" attraction between two oppositely charged objects. An AFM study of the force in the latter case has been published recently[9]. We emphasize that the attraction reported here is different than either phenomena. Unlike the first one, it takes place between two bare surfaces and unlike the latter it occurs also between identical surfaces. The resulting force is of a fundamentally shorter range compared with either phenomenon.

The force between a $5\,\mu m$ silica bead[10] glued to an AFM tip and an oxidized silicon wafer or the tip-mounted silica bead and an identical bead glued to a silicon wafer was measured using a commercial AFM[11]. The microscope was placed in an acoustic hood and the AFM's piezoelectric crystal was driven by a low noise synthesizer followed by a low pass filter[12]. This arrangement provided superior data compared with the commercial AFM electronics, especially at short distances. The commercial controller was only used to collect tip deflection signals and for reference measurements.

Force vs. distance curves measured in three different saline solutions, with the same Debye-Huckle (DH) screening length, $r_{DH} = 10\,nm$, are presented in Fig. 1. As evident from the inset, the force measured in the presence of sodium chloride decays exponentially at large distances with the expected decay length. Best fit to Poisson-Boltzmann (PB) theory yields for the potential at the edge of the diffuse layer and the dressed charge density, $\varphi_d \approx -50\,mV$ and $n_\sigma \equiv -\sigma/e \approx 0.025\,nm^{-2}$, respectively. These values agree with data published in the literature for silica surfaces[13]. When the sodium cation is replaced by a divalent magnesium cation the screening length shrinks considerably but repulsion persists. A dramatic change takes place when the sodium is replaced with cobalt hexamine, $Co(NH_3)_6^{+3}$ (CoH). In this case, the long range repulsion is suppressed and replaced by a short range exponential attraction. The decay length, $\approx 2\,nm$, is 5 times shorter than the screening length predicted by PB theory. This range of attraction is considerably shorter than any of the attraction lengths reported in the literature for the other attractive forces discussed in the introduction. Unlike van der Waals attraction the force decays exponentially with distance, clearly indicating the existence of a relevant length scale other than the screening length.

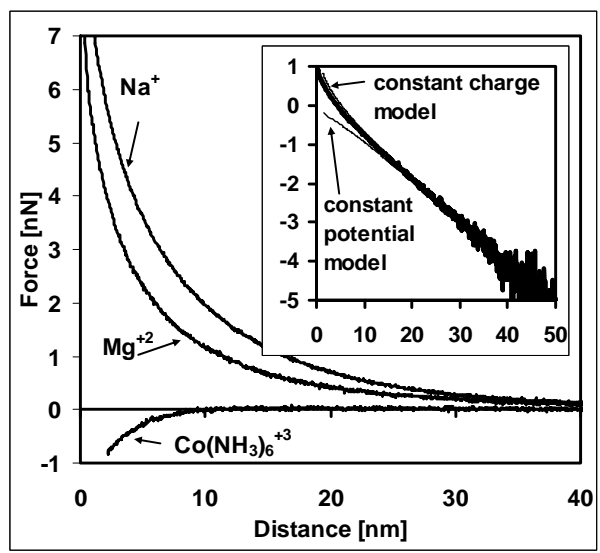

Fig. 1- Force vs. distance curves for three different cations and the same Cl⁻ anion. The ionic strength in all cases corresponds to $r_{DH} = 10\,nm$. The pH was set to 6 by negligible addition of HCl or NaOH. Inset- Semilog plot of the force vs. distance for the NaCl curve together with best fit to PB models with constant charge or constant surface potential.

Note the attraction curve is truncated at distances shorter than $\approx 2.5nm$. At that point the slope of the curve exceeds the cantilever spring constant, leading to instability and tip jump to the surface.

The evolution of attraction with increased $Co(NH_3)_6Cl_3$ concentration in a mixture with NaCl is depicted in Fig. 2a. The top curve corresponds to $\approx 0.93$mM pure NaCl. Substituting a minute quantity of sodium chloride with just $4.5 \times 10^{-6}$M $Co(NH_3)_6Cl_3$ (1:200 mixture) already leads to marked changes in the force curve. The repulsion is overall suppressed and an attractive component develops at small distances. Doubling the CoH concentration (1:100 curve) further suppresses repulsion and drives the maximal force point to longer distances. Finally, for a 1:25 mixture ($3 \times 10^{-5}$M $Co(NH_3)_6Cl_3$) the repulsion is almost completely suppressed and a pronounced, short range attraction appears. Note that even at this mixture ratio the ionic strength contributed by the CoH ions is marginal compared with that of sodium and chloride ions. The inset to Fig. 2a depicts the 1:25 curve at a magnified force scale. Clearly, weak repulsion persists down to $\approx 13nm$ where attraction commences abruptly.

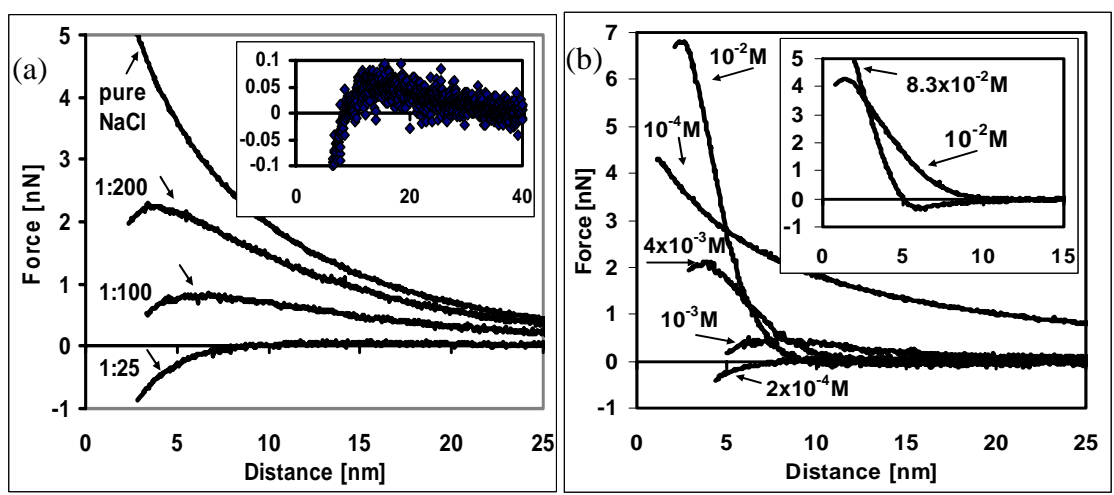

Fig. 2a- Force vs. distance curves for pure NaCl and three $Co(NH_3)_6Cl_3$ : NaCl mixtures. The ionic strength in all cases was set to give $r_{DH} = 10nm$. The pH was set to 6 by negligible titrations of either HCl or NaOH. Inset- a magnified view of the 1:25 mixture.

Fig. 2b- Force vs. distance curves for higher concentrations of $Co(NH_3)_6Cl_3$ (no NaCl). Inset- At very high $Co(NH_3)_6Cl_3$ concentrations (8.3x10$^{-2}$M curve) the short range repulsion is preceded by a small attraction (data taken with a different tip on a different sample than the main figure).

Interestingly, the attraction turns back into repulsion at higher CoH concentrations. The transition from repulsion to attraction and back to repulsion is demonstrated in Fig. 2b for pure $Co(NH_3)_6Cl_3$ solutions. At $10^{-4}M$ a long range repulsion between the surfaces dominates. Upon doubling the concentration the repulsion at large distances is suppressed and simultaneously a short range attraction appears. When the cobalt salt concentration is increased to $10^{-3}M$, the long range repulsion reappears, albeit with an attractive component at short distances. At even higher CoH concentrations the repulsion is further enhanced, still preserving an attractive component at short distances. The spatial range of repulsion shortens with CoH concentration, in line with the shrinking DH screening length. At even higher salt concentrations (inset to Fig. 2b), attraction appears below $\approx 8nm$ turning into repulsion at shorter distances.

The vanishing repulsion at intermediate CoH concentrations and its reappearance at higher concentrations are consistent with the well known charge reversal phenomenon[14], as well as with coagulation and resuspension phase diagrams of, e.g., DNA molecules in the presence of high valence cations[15,1]. It is also reflected in Zeta

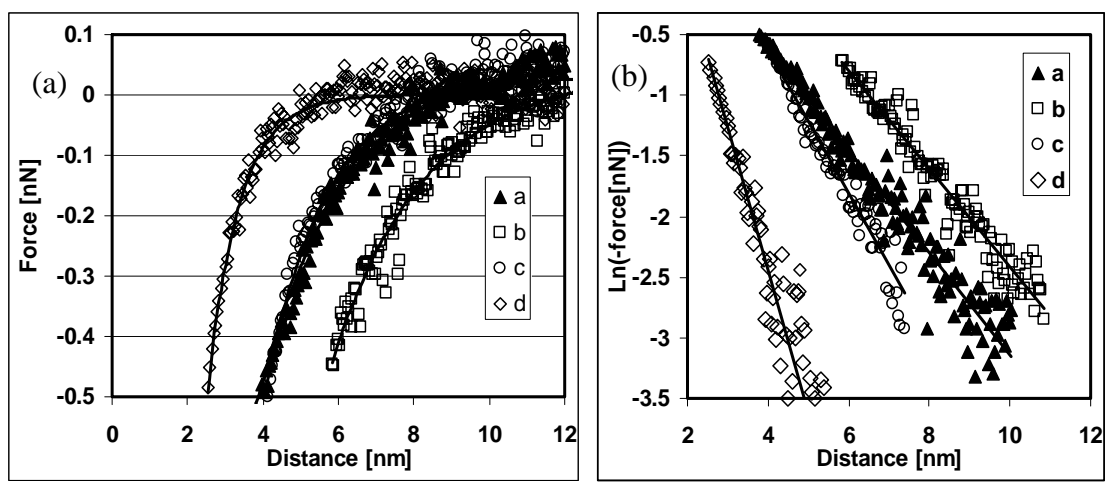

Fig. 3a- Four attraction curves fitted with single exponents. Curves a-d correspond respectively to: 0.75mM NaCl:0.03mM Co(NH$_3$)$_6$Cl$_3$ mixture, 0.1mM Co(NH$_3$)$_6$Cl$_3$, 0.2mM Co(NH$_3$)$_6$Cl$_3$, 23.75mM NaCl:0.1mM Co(NH$_3$)$_6$Cl$_3$ mixture. Solid lines represent best exponential fit. Fig. 3b- Semilog plots of the same data. The calculated DH screening lengths (a-d): 10nm, 12.6nm, 8.9nm, and 2nm. The measured decay lengths of attraction are (a-d): 2.4nm, 2.5nm, 1.7nm, and 0.85nm.

potential measurements we have carried out on the same type of colloids.

As seen in Figs. 3, the measured attractive force is well fitted, in all experiments, by a single exponent. Although the decay lengths of the force are considerably shorter than the corresponding calculated DH lengths (see values in the caption to Fig. 3b), the former are monotonic in the latter, namely, the shorter DH length is, the faster attraction decays with distance.

The net surface charge density and the dimensionless potential, $\psi_d \equiv e\varphi_d/T$, at the edge of the diffuse layer are extracted from force vs. distance curves by fitting the long distance exponential tails to a PB model corrected to a planar geometry by the Derjaguin approximation[16]. Close to neutrality (zero force at long distances), where attraction dominates, the potentials are small and the linear approximation to the PB equation suffices. Away from that point the linear approximation slightly overestimates $\psi_d$ but in the absence of an analytic PB theory for a mixture of different Z ions we continue to use it beyond its strict validity. The extracted potential is depicted in Fig. 4 for two NaCl:Co(NH$_3$)$_6$Cl$_3$ mixtures. The low and high NaCl curves correspond to $r_{DH} = 10 nm$ and $2 nm$, respectively. The measured screening lengths were significantly longer than the calculated values, especially near neutrality. The two high CoH points in the low NaCl curve were taken without NaCl since $r_{DH}$ was anyway shorter than $10 nm$ at these Co(NH$_3$)$_6$Cl$_3$ concentrations.

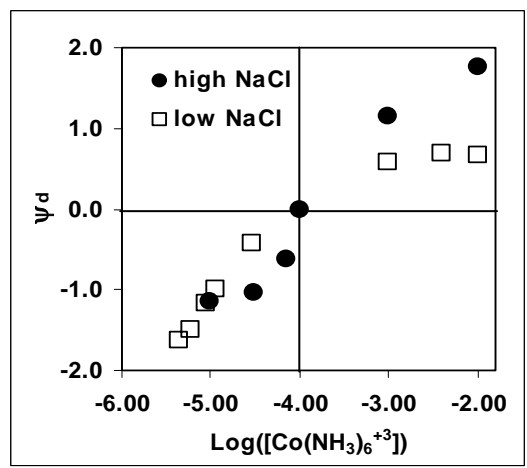

Fig. 4 – Extracted dimensionless potential at the edge of the diffusive layer vs. CoH concentration and two sodium chloride concentrations (see text).

Strikingly, the neutrality point, $\psi_d = 0$, $[\mathrm{Co(NH_3)_6^{+3}}] = 10^{-4} M$, is found to be independent of NaCl concentration. To account for this feature we assume that hydrogen, sodium, and CoH cations compete for the negatively charged silanol terminals to form $\mathrm{SiOH}$, $\mathrm{SiONa}$, or $\mathrm{SiOCo(NH_3)_6^{+2}}$ groups. Combining the laws of mass action for the three association reactions the surface charge density takes the form

$$\sigma = -en_0 \frac{1 - 2K_{Co}\left[Co(NH_3)_6^{+3}\right]_\infty \exp(-3\psi_0)}{1 + \left(K_H\left[H^+\right]_\infty + K_{Na}\left[Na^+\right]_\infty\right)\exp(-\psi_0) + K_{Co}\left[Co(NH_3)_6^{+3}\right]_\infty \exp(-3\psi_0)}. \quad (1)$$

Here, $K_H$, $K_{Na}$, and $K_{Co}$ stand for the respective association constants, $[\ ]_\infty$ are the corresponding cation concentrations away from the surfaces, $n_0$ is the density of ionizable silanol groups, and $\psi_0$ is the dimensionless potential at the surface. In principle the latter potential may be different for the three ions[17] due to their different size but this subtlety is neglected here. At neutrality, $\sigma$ and $\psi_0$ vanish leading to $K_{Co} = \left(2\left[Co(NH_3)_6^{+3}\right]_\infty\right)^{-1}$, **independent** of all other parameters. The crossing point of the two curves at neutrality thus follows naturally from Eq. (1). The same result also yields $K_{Co} = 5 \times 10^3 M^{-1}$ at $\left[Co(NH_3)_6^{+3}\right]_\infty = 10^{-4}M$. Analytical and numerical calculations[1,2,5,18] show that Coulomb correlations between the high valence ions amplify their condensation. In our language, the association "constant" should therefore grow with condensate density. It is worth mentioning that McLaughlin *et al.*[19] in their analysis of Zeta potential data to quantify divalent cation adsorption to membranes, and charge reversal by these cations, found an association constant which is **independent** of condensate concentration. Charge reversal may, hence, take place also in the absence of Coulomb correlations.

With the short range attraction in mind we turn to estimate the average surface density of the condensed CoH ions. The association constants of protons and sodium cations are estimated by measuring the force in the absence of cobalt hexamine chloride. Compilation of data taken at different pH values (not shown) gives $K_H \approx 10^3 M^{-1}$ and negligible $K_{Na}$. This value of $K_H$ is about 5 times smaller compared with ref. 20.

At neutrality, neglecting sodium ions, the protons and CoH coverages are given by
$$\alpha = K_H\left[H^+\right]_\infty \Big/ \left(1 + K_H\left[H^+\right]_\infty + K_{Co}\left[Co(NH_3)_6^{+3}\right]_\infty\right) = K_H\left[H^+\right]_\infty \Big/ \left(3/2 + K_H\left[H^+\right]_\infty\right)$$
and
$$\beta = K_{Co}\left[Co(NH_3)_6^{+3}\right]_\infty \Big/ \left(1 + K_H\left[H^+\right]_\infty + K_{Co}\left[Co(NH_3)_6^{+3}\right]_\infty\right) = \left(3 + 2K_H\left[H^+\right]_\infty\right)^{-1},$$
respectively. In particular at pH6, where most our data were taken, $\alpha \approx 10^{-3}$ and $\beta \approx 1/3$. With accepted densities of ionizable silanol groups, $n_0 \approx 0.3 \div 1 nm^{-2}$, the average distance between condensed CoH ions may thus vary between $R_{Co} = 3$ and $1.7 nm$. The corresponding classical plasma parameter of the condensate is given by $\Gamma \equiv Z^2 l_B \sqrt{\pi n_{Co}} \approx 3.5 \div 6$ where $\ell_B = 0.7 nm$ is the Bjerrum length. In a Wigner crystal type of model[18] the correlation contribution to the chemical potential hence comes at neutrality to $-\mu_{WC} \equiv k_B T \left(1.65\Gamma - 2.61\Gamma^{1/4} + 0.26 \ell n \Gamma + 1.95\right) \approx 4.5 \div 8 k_B T$, in accordance with the strong correlations needed for charge inversion.

We turn now to test the model presented in Ref. 4 according to which attraction results from interaction between Wigner crystals on the two plates. The only length scale in the model is the distance between condensed charges, $R_{Co}$. Explicit calculations show that the exact scale is $R_{Co}/2\pi$, less than $0.5 nm$ according to our estimates for the neutrality point. This scale is almost five times smaller than the experimental decay length of curve (a) in Figs. 3. There is no room for adjusting $R_{Co}$ to fit the exponential attraction since that would lead to $R_{Co} \approx 14 nm$, $\Gamma \approx 0.85$ and, hence, complete melting of the crystal by thermal fluctuations, leading to contradiction with the assumptions of the model. The simple model[4] is also independent of screening by other ions. As evident from comparison between curves (a) and (d) in Figs. 3, addition of sodium chloride reduces the attraction range from $2.4 nm$ to $0.85 nm$. The AFM data are hence **inconsistent** with the simple model of attraction between two phase-locked Wigner crystals.

The model presented in Ref. 6 does not account for the data either (albeit to a lesser degree). At distances longer than the DH screening length it predicts an attractive force decaying exponentially on a scale $r_{DH}/2$. This length is about 2 times longer compared with the experimental data at low NaCl concentrations (curve (a) in Fig.3) and only 15% longer than the decay length observed at high NaCl concentrations (curve (d), Fig. 3). The model, though, predicts a non exponential decay of the force at distances shorter than $r_{DH}$. This prediction disagrees with the observed exponential dependence.

In summary. Two nominally identical silica surfaces are found to attract each other in the presence of intermediate concentrations of three valent cations. The attraction decays exponentially with a characteristic length much shorter than the DH screening length. The measured force curves facilitate indispensable testing of models for the widely observed attraction between similarly charged objects in the presence of high valence ions.

Acknowledgement

We are grateful to B. I. Shklovskii for numerous discussions over the years. The work was supported by the Israeli Science Foundation.

**References**

[1] For recent reviews see: W. M. Gelbart *et al*, *Phys. Today*, **53** 38 (2000) and references therein, Y. Levine, *Rept. Prog. Phys.* **65** 1577 (2002) and references therein,

[2] For a recent review see A. Yu Grosberg, T. T. Nguyen, and B. I. Shklovskii, *Rev. Mod. Phys.* **74**, 329 (2002) and references therein.

[3] F. Oosawa, *Biopolymers* **6**, 1633 (1968)

[4] I. Rouzina and V. A. Bloomfield, *J. Phys. Chem* **100**, 9977 (1996)

[5] A. G. Moreira and R. R. Netz, *Europhys. Lett.* **57**, 911 (2002)

[6] R. Zhang and B. I. Shklovskii, *Phys. Rev.* **E 72**, 021405 (2005)

[7] Israelachvili, J. N. & Pashley, R. (1982) *Nature* **300**, 341–342.

[8] E. E. Meyer, Q. Lin, T. Hassenkam, E. Oroudjev, and J. N. Israelachvili, *PNAS* **102**, 6839 (2005) and references therein.

[9] K. Besteman, M. A.G. Zevenbergen, H. A. Heering, and S.G. Lemay, *Phys. Rev. Lett.* **93**, 170802-1 (2005)

[10] $5\mu m$ silica beads (Bangs Laboratories, Inc.) were glued to the AFM tip and the substrate using (Glass Bond, Loctite). After UV curing of the glue the tip and substrate were placed in oxygen plasma (Axic Multimode HF-8, 200 mTorr, 100 W) for 50 min and immediately introduced into the AFM chamber. In cases where the force was measured relative to a silicon wafer the latter was submitted to a similar oxygen plasma treatment.

[11] Multimode Nanoforce – Digital Instruments

[12] HP3325B synthesizer and a homemade low pass filter with a 6 Hz roll-off frequency.

[13] W. A. Ducker, T. J. Senden, and R. M. Pashley, *Langmuir* **8**, 1831 (1992)

[14] J. N. Israelachvili, *Intermolecular and Surface Forces, second edition*, Academic Press 1992, p. 237.

[15] M. Saminathan *et al.*, *Biochemistry* **38**, 3821 (1999)

[16] See, e.g., Ref. 14 p. 161.

[17] R.J. Hunter, *Foundations of Colloid Science*, Oxford University Press NY 2001, Ch. 10.

[18] B. I. Shklovskii, *Phys. Rev. Lett.* **82**, 3268 (1999); *Phys. Rev.* **E 60**, 5802 (1999)

[19] S. McLaughlin *et al.*, *J. Gen. Physiol.* **77**, 445 (1981)

[20] B. V. Zhmud, A. Meurk, and L. Bergström , *J. Colloid Interface Sci.* **207**, 332 (1998)